\documentclass[prl,showpacs,twocolumn]{revtex4}
\usepackage{amssymb}
\usepackage{amsmath}
\usepackage{graphicx}

\def    \bse{\begin{subequations}}
\def    \ese{\end{subequations}}

\def \be{\begin{equation}}
\def \ee{\end{equation}}
\def \bew{\begin{widetext}\begin{equation}}
\def \eew{\end{equation}\end{widetext}}
\def \bmlett{\begin{mathletters}}
\def \emlett{\end{mathletters}}

\def \tchi{\tilde{\chi}}

\def \hu{\hat{u}}

\def \ha{\hat{a}}

\def \hc{\hat{c}}
\def \hH{\hat{H}}
\def \hd{\hat{d}}

\def \omegam{\omega_M}
\def \omegadk{\omega_{\rm dk}}

\def \Nm{N_M}

\def    \neff{\bar{n}_{\rm h}}
\def  \ampdecay{ \lambda }

\def \hcd{\hat{c}_{\rm dk}}
\def \hdinmode{\hat{u}_{1} }

\def \bchi{\boldsymbol{\chi}}
\def \btchi{\boldsymbol{\tilde{\chi}}}

\begin{document}

\title{Using interference for high fidelity quantum state transfer in optomechanics}
\author{Ying-Dan Wang}
\affiliation{Department of Physics, McGill University, 3600 rue
University, Montreal, QC Canada H3A 2T8}
\author{Aashish~A.~Clerk}
\affiliation{Department of Physics, McGill University, 3600 rue
University, Montreal, QC Canada H3A 2T8}

\date{Oct. 20, 2011}

\begin{abstract}
We revisit the problem of using a mechanical resonator to perform
the transfer of a quantum state between two electromagnetic cavities
(e.g.~optical and microwave).  We show that this system possesses an
effective mechanical dark state which is immune to mechanical
dissipation; utilizing this feature allows highly efficient transfer
of intra-cavity states, as well as of itinerant photon states.  We
provide simple analytic expressions for the fidelity for
transferring both Gaussian and non-Gaussian states.
\end{abstract}


\maketitle




\textit{Introduction-- }The field of quantum optomechanics, where a
mechanical resonator is coupled to photons in a cavity, has seen
remarkable recent progress.  Milestones include using the backaction
of photons to cool a mechanical resonator to near its ground
state~\cite{Teufel2011b,Painter2011b}, and the observation of strong
coupling effects~\cite{Aspelmeyer2009,Teufel2011a,Kippenberg2011}.
The ability of a mechanical resonator to couple to diverse
electromagnetic cavities naturally leads to what is perhaps the most
promising application of this field: the possibility of efficiently
transferring a quantum state between photons with vastly differing
wavelengths~\cite{Tian2010,Regal2011,Safavi2011, Lukin2010}. Such
state transfer would have direct utility in quantum information
processing (e.g.~the transfer of quantum information from a
superconducting qubit in a microwave circuit QED setup to an optical
photon, or highly non-classical microwave states as prepared in
Ref.~\cite{Hofheinz2009} to optical photons).

\begin{figure}
\begin{center}
\includegraphics[bb=56 32 571 239,scale=0.46,clip]{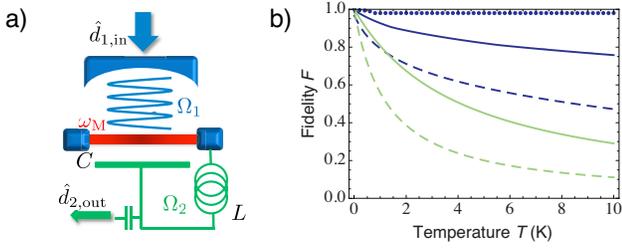}
\end{center}
\vspace{-0.5cm} \caption{(Color online) (a) The two-cavity
optomechanical system. (b) Fidelity of the double swap (DS) protocol
(light green) and the adiabatic transfer (AT)  protocol (blue) to
transfer a coherent state $|\alpha=1\rangle$, where $G/2\pi=2$~MHz
(solid), $G/2\pi=0.5$~MHz (dashed), and $\gamma=2 \pi \times 1$~KHz,
$\omega_M / 2 \pi=10$~MHz. Cavity 1 (2) is a microwave (optical)
cavity: $\Omega_1 / 2 \pi =10$~GHz ($\Omega_2 / 2 \pi = 100$~THz), $
\kappa_1 = \kappa_2 =  2 \pi  \times 50$~KHz. For DS, $G_1 = G_2 =
G$ and the total transfer time is $t^\text{s}=\pi/G$; while for AT,
we used an optimal modulation where $G_1^2(t) + G_2^2(t) = G^2$ is
constant~[28]. The blue dotted line corresponds to AT of $| \alpha =
0.1 \rangle$ with $G/2\pi=2$~MHz; as the amplitude decay effect is
negligible, a long transfer time can be used to suppress heating
caused by non-adiabatic transitions. For DS, the fidelity is mainly
limited by heating, hence $F$ versus $T$ for $|\alpha = 0.1\rangle$
is almost indistinguishable from $|\alpha = 1 \rangle$. }
\label{fig:lin_stirap}
\end{figure}

Previous investigations of this problem have largely considered
schemes based on two successive ``swap" operations in a two-cavity
optomechanical system (Fig.~\ref{fig:lin_stirap}a).  One pulses the
optomechanical interactions to first exchange the states of the
first cavity and the mechanical resonator; this is then repeated to
exchange the mechanical and the second cavity
states~\cite{Tian2010,Regal2011, Parkins1999}).
While intuitively simple, achieving high-fidelity with this protocol is only possible with
low levels of cavity and mechanical dissipation; \emph{we quantify
this below}.  In particular, one requires extremely low mechanical
bath temperatures.  This is true even if the mechanics is initially
prepared in its ground state~\cite{Tian2010,Regal2011}, as heating
during the transfer nonetheless degrades the state. Aspects of this
swap-scheme were recently implemented
experimentally~\cite{HailinWang2011}.

Given the above, it would be highly advantageous to find new state
transfer schemes less sensitive to mechanical dissipation. This is
the goal of this paper.  We show that the two-cavity optomechanical
system possesses a mode which is delocalized between the two
cavities while simultaneously being decoupled from the mechanical
dissipation; we term this decoupled mode a ``mechanically-dark"
mode, as it is analogous to an atomic state which is protected
against optical excitation by destructive interference
\cite{Fleischhauer2005}.  We show that by using this dark mode, one
can perform high-fidelity quantum state transfer of intra-cavity
states at levels of mechanical dissipation where the conventional
double-swap scheme is essentially unusable.

We also show that this dark mode can be used for efficient transfer
of \textit{itinerant} photons (e.g.~transferring the state of
photons incident on a microwave cavity to the state of photons
leaving an optical cavity). This approach is particularly
attractive, as it does not require any time-dependent variation of
optomechanical couplings. Further, if one is willing to only
consider the transfer of small-bandwidth states, the scheme can also
be used without requiring optomechanical strong coupling.  We
quantify analytically the fidelity of this scheme for Gaussian
states (in a way that allows easy comparison against the
intra-cavity transfer schemes mentioned above), as well as
non-classical states; we also consider limitations on the bandwidth
of the states that can be transferred.  These analytic expressions
yield a simple intuitive picture of the factors limiting fidelity.
In the limit of weak coupling, this itinerant-photon transfer scheme
is equivalent to that described by Safavi-Naeni et
al.~\cite{Safavi2011} (though that work did not discuss fidelities,
strong coupling, or the role of the dark mode).


\textit{Model-- } We consider an optomechanical system where a
single mechanical resonator is simultaneously coupled to both an
optical cavity and a microwave cavity via dispersive couplings (see
Fig.~\ref{fig:lin_stirap}a).  We also focus on the standard
situation where the bare optomechanical coupling is enhanced by
strongly driving each cavity, resulting in effective linear
couplings (see, e.g.,~\cite{Marquardt07,WilsonRae07}).  We work in
an interaction picture with respect to the two cavity drives, and in
a displacement picture with respect to the average (classical) field
in each cavity.  The Hamiltonian is:
\begin{equation}
    \hH = \omegam \ha^\dag \ha - \sum_{i=1,2}
    \left[
         \Delta_i \hd^\dag_i \hd_i
        -  G_{i} \left( \ha^\dag \hd_i + \hd^\dag_i \ha   \right)
    \right]
        + \hH_{\rm diss}      \label{ham}
\end{equation}
Here, $\omegam$ ($\ha$) is the mechanical frequency (annihilation
operator), $\hd_i$ is the annihilation operator of cavity $i$
($i=1,2$) in the displaced frame, and $\Delta_i$ is the detuning of
the drive applied to cavity $i$.  The driven optomechanical coupling
between the mechanical resonator and cavity $i$ is denoted as $G_i$;
note that these are proportional to the drive amplitude applied to
cavity $i$, and thus can be controlled in time. $H_{\rm diss}$
describes the damping and driving of the two cavities and mechanical
resonator by independent Ohmic baths.  We let $\gamma$ ($\kappa_i$)
denote the damping rate of the mechanical resonator (cavity $i$),
and let $\Nm$ ($N_i$) denote the corresponding bath temperature
(expressed as a number of thermal quanta). We also assumed the
optimal situation where each cavity is far into the
resolved-sideband regime $\omegam \gg \kappa_i$, and where each
cavity is driven near the red-detuned mechanical sideband
(i.e.~$\Delta_i \sim - \omegam$).  This permits us to make a
rotating wave approximation in writing the optomechanical
interactions, resulting in a ``beam-splitter" form which is optimal
for state transfer~\cite{Parkins1999}.

\textit{Double-swap protocol-- }The optomechanical interactions in Eq.~(\ref{ham})
can be used to swap states between the three modes of the system \cite{Parkins1999, Tian2010, Regal2011}.
The swap protocol involves first turning on the interaction $G_1$ for a time $t_{1}^{\text{s}} = \pi / (2 G_1)$ (while $G_2 = 0$),
which if $\gamma = \kappa_1 = \kappa_2 = 0$
would swap the states of cavity $1$ and the mechanical resonator (i.e.~$\hat{a}(t_{i}^{\text{s}})=-i\hat{d}_{i}(0)$ and $\hat{d}_{i}(t_{i}^{\text{s}})=-i\hat{a}(0)$).  One then shuts off $G_1$ and turns on $G_2$ for a time $t_{2}^{\text{s}}$ to swap the mechanical state to cavity 2.

The presence of both mechanical and cavity dissipation degrades the fidelity of this protocol.  To quantify this,
we consider the simple case of transferring a Gaussian state, and calculate the Uhlmann fidelity $F$  \cite{Uhlmann1976}
between initial and final states.  Letting $\rho_1$ ($\rho_2$) denote the density matrix of cavity 1(2) at the start (end) of the transfer, we find
\footnote{An expression for the fidelity of Gaussian state transfer was given in Ref.~\cite{Wallquist2010}; however,
this work neglected the effects of cavity and mechanical damping, and hence their expression lacks the exponential factor in our Eq.~(\ref{fg}).}:
\begin{equation}
    F  \equiv
        \left( \mathrm{Tr}[(\sqrt{\rho _{1}}\rho_{2}\sqrt{\rho _{1}})^{1/2}]\right) ^{2}
    =
    \frac{1}{1 + \neff }\exp \left( -\frac{ \ampdecay^{2}}{ 1 + \neff }\right).
    \label{fg}
\end{equation}
Note that we will optimize the fidelity over simple rotations in
phase space (so that if $\rho_2$ is a rotated version of $\rho_1$,
$F=1$). $F$ depends on just two parameters: $\neff$ represents the
heating of the state during the protocol by noise emanating from
cavity and mechanical dissipative baths, while $\ampdecay$
characterizes the decay of the mean value of $\hd$ due to cavity and
mechanical damping.  Efficient transfer requires minimizing both
these effects.  In the double-swap protocol, the amplitude-decay
will completely suppress $F$ unless one is in the strong coupling
limit $G_i > \kappa_i$. In this relevant limit, and for the case
where the state to be transferred is a coherent state $|\alpha
\rangle$, we find the simple result \footnote{The full expression
for $F$ for an arbitrary Gaussian state is rather unwieldy, and will
be presented elsewhere~\cite{Wangprep}.}:
\begin{equation}
    \neff = \sum_{i}\frac{\gamma\Nm+
        \kappa_{i}N_{i}}{2}t_{i}^{\text{s}}
    \text{, }
        \ampdecay = \left| \alpha \right|
    \sum_{i}\frac{\kappa _{i}+\gamma}{4}t_{i}^{\text{s}},
\label{ds}
\end{equation}
where $\left( \gamma\Nm +\kappa N_{i}\right) /2$ is the average
heating rate and $(\kappa _{i}+\gamma)/4$ is the average amplitude
decay rate during each time interval.   We have assumed the optimal
situation where the mechanical resonator is initially in its ground
state (\cite{Tian2010,Regal2011}). Despite this pre-cooling, the
mechanical contribution to $\neff$ can still be large.
%
%
One thus requires an extremely low mechanical bath temperature to
ensure good fidelity using swap scheme (see
Fig.~\ref{fig:lin_stirap}b).

\textit{Effective mechanically-dark mode-- }
From Eq.~(\ref{ds}), we see that the heating $\neff$ due to mechanical noise in the double-swap
scheme is simply the heating rate times transfer time, and hence scales as
$1/G$.  We now show that transfer protocols exist where the mechanical heating effect
is even more greatly suppressed with increasing $G$.  This is possible by
making use of a mode of the two-cavity
optomechanical system which is simultaneously delocalized between both cavities, but at the same time
is largely immune to to mechanical dissipation.

Focusing as before on the case where each cavity is driven at the
red-detuned sideband ($\Delta_i = - \omegam$), we first note that
the coherent part of the system Hamiltonian $\hH_0 = \hH - \hH_{\rm
diss}$ can be diagonalized as $\hH_0 = \sum_{j} \hbar \omega_j
\hc_j^\dag \hc_j$ with $j={\pm,0}$.
$\hc_{\pm} =
        \left(  \left( 2(  G_1^2 + G_2^2) \right)^{-1/2}  \left(G_1 \hd_1 + G_2 \hd_2 \right)
        \pm   \ha /  \sqrt{2}   \right)$
describe hybridized mechanical and cavity  modes
with frequencies $\omegam \pm \sqrt{G_{1}^{2}+G_{2}^{2}}$, whereas
\begin{eqnarray}
    \hat{c}_0\equiv \hcd & = &
        \left(  G_1^2 + G_2^2 \right)^{-1/2}  \left(-G_2 \hd_1 + G_1 \hd_2 \right)
\end{eqnarray}
describes a combination of cavity modes which is decoupled from the mechanics.
We thus refer
to $\hcd$ as a ``mechanically dark" mode.  Note that its frequency is $\omegadk = \omegam$, independent of coupling.
As we now demonstrate, utilizing this mode allows the efficient transfer of both intra-cavity
and itinerant photon states.

\textit{Adiabatic transfer-- } Consider first the same problem
addressed by the double-swap scheme, the transfer of an intra-cavity
state initially in cavity 1 to cavity 2.  This can be accomplished
by using an adiabatic passage approach, similar to the well-known
STIRAP scheme \cite{Bergmann1998}. One modulates $G_1(t)$ and
$G_2(t)$ so that the dark mode adiabatically evolves from being
$-\hd_1$ at $t=0$ to $\hd_2$ at the end of the protocol at a time
$t=t_f$.  The cavity state is thus transferred from cavity 1 to
cavity 2 using the the coupling to the mechanics, but without
actually populating the mechanics; the result is a greatly enhanced
protection against mechanical sources of dissipation.

Fig.~\ref{fig:lin_stirap}b shows how such an adiabatic transfer
protocol improves the state transfer fidelity over the double-swap
scheme when the mechanical heating effect is non-negligible.  When
transferring a Gaussian state, $F$ again takes the general form
described by Eq.~(\ref{fg}). The adiabatic ``dark state" transfer
protocol dramatically suppresses $\neff$ compared to the swap
scheme.  However, to remain adiabatic, the transfer must ideally
occur over a time long compared to $1/G$. Thus, similar to the swap
scheme, one needs strong coupling (i.e.~$\kappa_i \ll G_i$) to avoid
the amplitude-decay suppression of $F$ described by $\lambda$ in
Eq.~(\ref{fg}). Nonetheless, the greater resilience against
mechanical noise presents a strong advantage over the double-swap
scheme. Note that for the adiabatic-transfer fidelities in
Fig.~\ref{fig:lin_stirap}b, we take for each temperature an optimal
transfer time which represents a trade-off between heating (via
non-adiabatic transitions) and amplitude decay. A somewhat related
scheme for transferring atomic motional states was discussed in
Ref.~\cite{Parkins1999}; the uni-directional ``cascaded" coupling
used there is fundamentally different from that considered here.

%

\textit{Itinerant state transfer--}
While the previously discussed transfer schemes require a strong optomechanical coupling (i.e.~$G_i \gg \kappa_i$),
mechanically-mediated transfer is also possible in the opposite regime if the goal is to
transfer a narrow-bandwidth state of photons incident on cavity 1 to the state of photons leaving cavity 2
\cite{Safavi2011, Braunstein2003}.
%
We now show that the mechanically-dark state discussed above plays an important role in this itinerant-photon state transfer,
and even allows it to be highly effective in regimes of strong optomechanical coupling.
We begin by writing the Heisenberg-Langevin equations for our system \cite{Gardiner00, ClerkRMP}:
\begin{eqnarray}
\dot{\hat{a}} &=&-i\omega _{M}\hat{a}-\gamma\hat{a}-i\sum
G_{i}\hat{d}_{i}-\sqrt{2\gamma}\hat{a}_{\text{in}}  \notag \\
\dot{\hat{d}}_{i} &=&-i\Delta _{i}\hat{d}_{i}-\kappa
_{i}\hat{d}_{i}-iG_i\hat{a}-\sqrt{2\kappa _{i}}\hat{d}_{i,\text{in}}
\end{eqnarray}%
with $\hat{a}_{\text{in}}$ and $\hat{d}_{i,\text{in}}$ representing both input noise (taken to be white) and signals driving each
resonator.  Solving this equation in frequency domain yields
$\hat{\mathbf{A}} [\omega] =
    \bchi [\omega]\hat{\mathbf{A}}_{\text{in}}[\omega] $,
where $\bchi[\omega]$ is the $3 \times 3$ system susceptibility matrix
and $\hat{\mathbf{A}}=\{\hat{d}_{1}[\omega],\hat{d}_{2}[\omega], \ha[\omega ]\}$.  Using standard input-output theory \cite{Gardiner00},
 $\bchi$ determines the scattering matrix $\mathbf{s}[\omega]$ which relates output and input fields via
$\hat{\mathbf{A}}_{\text{out}}[\omega] =
    \mathbf{s}[\omega]\hat{\mathbf{A}}_{\text{in}}[\omega] $.
It will also be useful to transform the susceptibility using the $({\rm dk},+,-)$ basis to describe the cavity modes; we denote this matrix $\btchi$.

High fidelity transfer from $\hd_{1, \rm{in}}$ to $\hd_{2, \rm{out}}$ requires that over the input signal bandwidth,
the transmission coefficient $|s_{21}[\omega]|^2 \sim 1$, as well as that $|s_{23}[\omega]|^2 \sim 0$
(i.e.~negligible transmission of mechanical noise).
To quantify this, we consider a Gaussian input state in a temporal mode defined by
$   \hdinmode =
        \left(2 \pi\right)^{-1/2}
         \int d\omega f\left[ \omega \right]
\hat{d}_{1,\text{in}}\left[ \omega \right]$ (see,
e.g.,~Ref.~\cite{ClerkRMP}).  $f[\omega]$ describes a wavepacket
incident on cavity 1 which is localized in both frequency and time;
$\int d \omega | f[\omega] |^2 = 1$ to ensure that $\hdinmode$ is a
canonical bosonic destruction operator. The fidelity of transferring
this itinerant Gaussian state again takes the general form of
Eq.~(\ref{fg}), and the parameters $\neff$ and $\ampdecay$ can be
calculated analytically~\cite{Wangprep}.  For a coherent state input
$| \psi_{\rm in} \rangle \propto \exp\left(\alpha
\hdinmode^\dag\right) |0\rangle$:
\begin{eqnarray}
    \neff &=&
        \sum_{i=1,2,M}\int d\omega
            \left\vert
                f \left[ \omega \right] s_{2i}\left[ \omega \right]
            \right\vert^{2}N_i
    ~~~ \\
    \ampdecay  &=&
        |\alpha| \max_{\tau} \left(
            1- \left|
                \int \, d\omega   e^{-i\omega \tau } s_{21}\left[\omega \right]
                \left| f \left[ \omega \right] \right|^2 \right|
        \right)  ~~~~~
\end{eqnarray}
We have optimized the final state $\rho_2$ in Eq.~(\ref{fg}) over a
time-translation $\tau$, so that if the output pulse is simply a
time-delayed copy of the input pulse, $F=1$.

To see how the mechanically-dark state aids itinerant state
transfer, consider first the simple case where the input state has a
narrow bandwidth. To be protected against mechanical dissipation,
one would ideally like the input state incident on cavity 1 to only
drive the dark mode of the two-cavity optomechanical system, i.e.
$\tchi_{\text{dk},1} \sim 0$.  Without dissipation, the dark mode is
energetically separated from the coupled modes $\hd_+, \hd_-$, and
hence this condition is achieved by using an input signal with mean
frequency $\omegam$.  Even in the presence of dissipation (and the
consequent lifetime broadening of mode energies), one still obtains
$\tchi_{\pm,1}[\omega] /  \tchi_{\text{dk},1}[\omega] \propto
1/\sqrt{C_1 C_2}$, where the co-operativity $C_i =  G_i^2 / \gamma
\kappa_i $. As long as the co-operativity parameters $C_i \gg 1$,
the input signal only appreciably excites the mechanically-dark
mode. This is similar to optomechanical electromagnetic-induced
transparency \cite{Agarwal2010,Kippenberg2010,Painter2011a}:  an
analogous interference prevents the coupled modes $\hd_+, \hd_-$
from being excited.

Good fidelity also requires that the dark state, once excited
by the input state, only leaks out via cavity 2.  This requires a
destructive interference between the promptly reflected input signal
and the wave leaving the dark mode via cavity 1.  For $C_1, C_2 \gg
1$, this interference cancellation results in the simple impedance
matching condition $C_1 = C_2 \equiv C$, i.e.:
\begin{equation}
    G_1^2 / \kappa_1 = G_2^2 / \kappa_2
    \label{con}
\end{equation}%

\begin{figure}[tp]
\begin{center}
\includegraphics[width= 0.99 \columnwidth]{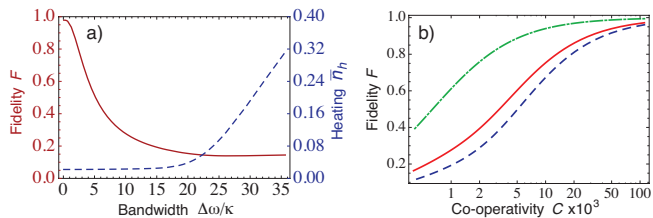}
\end{center}
\vspace{-0.5cm} \caption{Fidelity for transferring itinerant photon
states at $T = 2$~K. (a) Fidelity (red, solid) and heating (blue,
dashed) versus input bandwidth for transferring $|\alpha=
\sqrt{3}\rangle$ coherent state. (b) Fidelity versus co-operatively
$C = G^2 / \kappa \gamma$
in the narrow-bandwidth limit. The input states are $|\alpha=
\sqrt{3} \rangle$ (green, dash-dot), $|n=3\rangle$ Fock state
(blue, dashed), and $(|1\rangle+|3\rangle)/\sqrt{2}$ (red, solid).
Unless specified here, the system parameters are the same as the
dashed line in Fig.~\ref{fig:lin_stirap}b.} \label{fig:incident}
\end{figure}

Taking our input mode $|f[\omega]|^2$ to have mean frequency
$\omegam$ and a Gaussian profile with variance $\Delta \omega^2$, and
assuming $C_1 = C_2 \gg 1$, we find to leading order in $\Delta
\omega$:
\begin{eqnarray}
    \neff  &\approx &
        \frac{\Nm}{4C}~
            \left( 1+\left( \frac{\Delta \omega }{G}\right)^{2}\left( 1-\frac{\kappa ^{2}}{16G^{2}}\right) \right) ~
            \label{eq:itinerantneff}\\
    \ampdecay  &\approx & | \alpha | \left[
            \frac{1}{8C}+\left(
            \frac{2 \Delta \omega }{\kappa }\right) ^{2}\left( 1+\left(
            \frac{\kappa ^{2}}{8G^{2} }\right) ^{2}\right)
        \right]
        \label{eq:itinerantampdecay}
\end{eqnarray}
Good fidelity requires a high co-operativity $C \gg
|\alpha|, \Nm$. In the weak-coupling regime $G < \kappa$, one also needs
$\sqrt{| \alpha |} \Delta \omega \ll (G^2 / \kappa) $, which
reflects the width of the $s_{21}[\omega]$ transmission resonance.
In the opposite regime $G \gg \kappa$,  one needs $\Delta \omega
\leq \kappa / \sqrt{| \alpha |}$ as shown in
Fig.~\ref{fig:incident}a. Further, we see that in comparison against
the double-swap scheme, the mechanical-heating effect described by
$\neff$ is reduced by a large factor $\kappa / G$.  The expression
of $\neff$ is the usual {\it weak-coupling} expression for the
mechanical temperature cavity cooling
\cite{Marquardt07,WilsonRae07}; unlike cavity-cooling, it describes
$\neff$ in both weak and strong coupling regimes.

\textit{Transfer of non-classical itinerant states-- } Given the
advantages of the itinerant transfer scheme, it is also interesting
to consider how well it is able to transfer non-classical states.
While in general it is difficult to obtain analytic expressions for
the evolution of non-Gaussian states, we show that here, one can obtain
useful and reliable analytic approximations.

We again consider an input mode in a given temporal mode $\hu_1$; we
take this mode to be centered on $\omegam$, and for simplicity, to
have a vanishingly small bandwidth $\Delta \omega$.  Suppose now the
input state incident on cavity 1 is a Fock state of this mode $| n
\rangle \propto \left(\hu^\dag_{1} \right)^n | 0 \rangle$. We also
take the noise driving both cavities to be zero-temperature ($N_1 =
N_2 = 0$), but allow the mechanical resonator to be driven by
thermal noise.
Letting  $p_{\rm th}(q, N_m)$ be the probability of having $q$ thermal quanta incident on the mechanical resonator,
the fidelity can be decomposed as
\begin{equation}
    F = \sum_{r=0}^\infty P(r , n) = \sum_{r=0}^\infty \sum_{q=0}^{\infty }
        p_{\rm th}(q, N_m)
        \left\vert f_{q}^{\left( r,n\right) }\right\vert ^{2}  \label{fock}
\end{equation}
where $P\left(r ,n \right)$ is the probability of having $r$ outgoing photons leaving cavity 1 and $n$ photons leaving cavity 2, and
\begin{eqnarray}
            &&
    f_{q}^{\left( r,n\right) } =
        \sqrt{\left(
            \begin{array}{c}
                q \\
                r
            \end{array}%
        \right)} \,
                \left( s_{21}\right) ^{n}\left( s_{33}\right) ^{q}
    \sum_{j=0}^{r}
        \left(
            \frac{s_{13}}{s_{33}}\right) ^{r}
        \left(
            \begin{array}{c}
                n \\
                j%
        \end{array}
        \right)
        \left(
            \begin{array}{c}
                r \\
                j%
        \end{array}
        \right)
        \nonumber
             \\
    &&
        \label{eq:fqrn}
        \left(
        \frac{s_{11}s_{23}} { s_{21}  s_{13} }\right)^{j}{}_{2}
            F_{1}\left(j-n,r-q;1+j;\frac{s_{31}s_{23}}{s_{21}s_{33}}\right)
\end{eqnarray}
where $_{2}F_{1}$ is the hypergeometric function and
$\mathbf{s}\equiv \mathbf{s}\left( \omega _{m}\right) $.

Note that in the regime of optimal state transfer $C_1 = C_2 \equiv C \gg 1$,
the probability of having photons leave cavity 1 is small:  the dark state effectively prevents mechanical
photons from contributing, and Eq.~(\ref{con}) ensures minimal reflection of signal photons.  One can thus get a good approximation
by simply retaining the $r=0$ and $r=1$ term in Eq.~(\ref{fock}):  $F$ is approximately just the probability of obtaining $n$ photons in the cavity 2 output mode
and at most one photon leaving cavity 1.  This
is a rigorous lower bound on the exact fidelity, and is exact to order $1/C$.


In the limit $C \gg 1$, one finds that to leading order in $1/C$ the
fidelity for transferring the $n$-photon itinerant Fock state is $
    F \simeq 1 -  \left[ \Nm \left(3 + 2n \right) + n
    \right]/4C$.
For $\Nm \gg 1$, the condition for a near-unity fidelity is thus $C \gg \Nm n$; for a large-$n$ Fock state, this is more stringent than the condition
for having a large fidelity transfer of a coherent state with $|\alpha| \sim \sqrt{n}$ (c.f.~Eqs.~(\ref{eq:itinerantneff}),(\ref{eq:itinerantampdecay})).
%

Finally, we note that the same approach can be used to compute the
fidelity of transferring an arbitrary pure input state of the form
$\left\vert \Psi _{1}\right\rangle =\sum_{m}c_{m}\left\vert
m\right\rangle $; the full expression is provided elsewhere~\footnote{See Supplemental Material for
details of the adiabatic pulse shape, and the transfer fidelity of
superposition Fock states.}.
The transfer fidelity of different non-Gaussian states together with
a coherent state (for realistic parameters) are shown in
Fig.~\ref{fig:incident}b.

\textit{Conclusions-- } In this paper, we have proposed using a
mechanically dark delocalized mode in a two-cavity optomechanical
system for quantum state transfer. We have demonstrated that both
intra-cavity states and itinerant photon states can be transferred
with high fidelity, using parameters within reach of current
experiments.

We thank S. Chesi, K. Lehnert,  O. Painter, C. Regal and A. Safavi-Naeini for useful
discussions.  This work was supported by the DARPA ORCHID program
under a grant from the AFOSR.


\end{document}


\title{Supplementary material for ``Using interference for high fidelity
quantum state transfer in optomechanics"}
\author{Ying-Dan Wang}
\affiliation{Department of Physics, McGill University, 3600 rue University, Montreal, QC
Canada H3A 2T8}
\author{Aashish~A.~Clerk}
\affiliation{Department of Physics, McGill University, 3600 rue University, Montreal, QC
Canada H3A 2T8}
\maketitle

\section{The pulse modulation used for adiabatic transfer}

The optimal pulse modulation for adiabatic transfer is one that keeps the
energy splitting between levels constant~\cite{Vasilev2009}; it thus
satisfies:
\begin{equation}
G_{1}^{2}\left( t\right) +G_{2}^{2}\left( t\right) =G^{2}
\end{equation}
In our paper, we used the following shape
\begin{equation}
G_{1}\left( t\right) =G\sqrt{\tanh \beta t},G_{2}\left( t\right)
=G\sqrt{1-\tanh \beta t}
\end{equation}
The parameter $\beta $ represents the speed of the modulation.
Slower modulation guarantees a better adiabaticity of the state
transfer, and hence better protection from the mechanical thermal
noise. However, cavity damping will destroy the state if the
duration of the state transfer is longer than the cavity decay time.
These two conflicting requirements results in an optimal, non-zero
value of $\beta $. High fidelity transfer requires $G\gg \beta \gg
\kappa $. The optimal value of $\beta$ also depends on the
temperature as the demand for a better adiabaticity rises as thermal
noise gets stronger. The red curves in Fig. 1b are obtained with
both the modulation rate $\beta $ and the total transfer time
$t^{\text{s}}$ optimized for each temperature.

\section{Superposition of Fock states}

We consider a superposition state as the input state of cavity 1
\begin{equation}
\rho _{in,1}=\sum_{mn}c_{n}^{\ast }c_{m}\left\vert m\right\rangle
_{1,\text{in}}\left\langle n\right\vert
\end{equation}%
$\left\vert m\right\rangle _{1,\text{in}}=\left( 1/\sqrt{m!}\right)
\left( \hat{d}_{1\text{,in}}^{\dag }\right) ^{m}\left\vert
0\right\rangle $ is the Fock state of the cavity-1 input, $c_{m}$ is
the corresponding amplitude. Here we have assumed a vanishingly
small bandwidth so that $\hat{u}_{1}\sim \hat{d}_{1,\text{in}}$.

The target state is the same superposition state of cavity 2 output
\begin{equation}
\rho _{\text{target}}=\sum_{mn}c_{n}^{\ast }c_{m}\left\vert m\right\rangle
_{2\text{,out}}\left\langle n\right\vert
\end{equation}%
with $\left\vert m\right\rangle _{2,\text{out}}=\left(
1/\sqrt{m!}\right) \left( \hat{d}_{2\text{,out}}^{\dag }\right)
^{m}\left\vert 0\right\rangle $) the Fock state of the cavity-2
output.

The transfer fidelity of such superposition state reads%
\begin{equation}
F=\sum_{q,r=0}^{\infty }p_{q}\sum_{mn}\sum_{d=-n}^{q-r}c_{m+d}^{\ast
}c_{n+d}c_{n}^{\ast }c_{m}f_{m,0,q}^{r,m+d,q-r-d}\left(
f_{n,0,q}^{r,n+d,q-r-d}\right) ^{\ast },
\end{equation}%
with%
\begin{eqnarray}
f_{n,0,q}^{r,n+d,q-r-d} &=&\sqrt{\frac{r!n!q!\left( n+d\right) !}{\left(
q-r-d\right) !}}\left( s_{21}\right) ^{n}\left( s_{33}\right) ^{q}\left(
\frac{s_{13}}{s_{33}}\right) ^{r}\left( \frac{s_{23}}{s_{33}}\right) ^{d}
\notag \\
&&\sum_{j=0}^{\min \left[ n,r\right] }\frac{_{2}F_{1}\left(
j-n,d+r-q;1+d+j;\frac{s_{31}s_{23}}{s_{33}s_{21}}\right) }{j!\left(
j+d\right) !\left( n-j\right) !\left( r-j\right) !}\left(
\frac{s_{11}s_{23}}{s_{21}s_{13}}\right) ^{k},
\end{eqnarray}%
where $_{2}F_{1}$ is the hypergeometric function.
$f_{n,0,q}^{r,n+d,q-r-d}$ is the amplitude for an input state
$\left\vert n,0,q\right\rangle _{\text{in}}$ scattering into an
output $\left\vert r,n+d,q-r-d\right\rangle _{\text{out}}$.
$\left\vert n,0,q\right\rangle $ denotes the state with $n$ photon
in cavity 1, $0$ photon in cavity 2, and $q$ phonon in the
mechanics. Note that the scattering matrix conserves the total
excitation, therefore the amplitudes between states with different
excitation numbers vanish.
